\documentclass[a4paper,aps,prl,notitlepage,superscriptaddress,longbibliography,twocolumn,10pt]{revtex4-2}

\usepackage{amsmath, amssymb, amscd, amsthm, amsfonts, physics}
\usepackage[utf8]{inputenc}
\usepackage{hyperref}
\usepackage[dvipsnames]{xcolor}

\usepackage{graphics,graphicx,float,xfp}
\usepackage{braket}
\usepackage{epstopdf}

\usepackage{pgfplots}
\pgfplotsset{compat=1.17}
\usepackage[normalem]{ulem}
\usepackage[compat=1.1.0]{tikz-feynman}

\usetikzlibrary{calc,patterns,angles,quotes,math,decorations.text,arrows.meta,decorations.markings}
\tikzstyle{world line}=[blue!70!black,line width=0.4]
\tikzstyle{world line thick}=[blue!80!black,line width=3]
\tikzstyle{world line t}=[purple!60,line width=0.4]

\tikzset{declare function={kruskal(\x,\c)  = {\fpeval{asin( \c*sin(2*\x) )*2/pi}};
}}
\tikzfeynmanset{warn luatex=false}

\begin{document}
\title{Finite complexity of the de Sitter vacuum}

\author{Suddhasattwa Brahma}
\email{suddhasattwa.brahma@gmail.com}
\affiliation{Higgs Centre for Theoretical Physics, School of Physics and Astronomy, University of Edinburgh, Edinburgh EH9 3FD, UK}

\author{Lucas Hackl}
\email{lucas.hackl@unimelb.edu.au}
\affiliation{School of Mathematics and Statistics, The University of Melbourne, Parkville, VIC 3010, Australia}
\affiliation{School of Physics, The University of Melbourne, Parkville, VIC 3010, Australia}

\author{Moatasem Hassan}
\email{moatasem1508@yahoo.com}
\affiliation{Higgs Centre for Theoretical Physics, School of Physics and Astronomy, University of Edinburgh, Edinburgh EH9 3FD, UK}

\author{Xiancong Luo}
\email{x.luo-35@sms.ed.ac.uk}
\affiliation{Higgs Centre for Theoretical Physics, School of Physics and Astronomy, University of Edinburgh, Edinburgh EH9 3FD, UK}

\begin{abstract}
The ER=EPR conjecture states that quantum entanglement between boundary degrees of freedom leads to the emergence of bulk spacetime itself. Although this has been tested extensively in String Theory for asymptotically anti-de Sitter spacetimes, its implications for an accelerating universe, such as our own, remain less explored. Assuming a cosmic version of ER=EPR for de Sitter space, we explore computational complexity corresponding to long-range entanglement responsible for bulk states on spacelike hypersurfaces. Rather remarkably, we find that the complexity (per unit volume) of the Euclidean vacuum, as an entangled state over two boundary CFT vacua, is {\bf finite} both in the UV and the IR, which provides additional evidence for {\it cosmic ER=EPR}. Our result seems to be a universal feature of spacetimes with horizons and, moreover, hints at new features of the thermofield double state for studying thermalization of any quantum system.
\end{abstract}

\maketitle

\noindent {\bf Introduction:} The Bekenstein-Hawking formula~\cite{Bekenstein:1972tm,Hawking:1975vcx}, relating the area of the event horizon of a black hole (BH) to its entropy, paved the way to our best understanding of quantum gravity, namely \textit{Holography}~\cite{tHooft:1993dmi,Susskind:1994vu,Maldacena:1997re,Bousso:2002ju}. It has revealed to us deep insights into non-perturbative quantum gravity by equating gauge theories that are ``dual'' to gravity in one higher dimension. Perhaps the most remarkable of them being that quantum entanglement between localized regions of spacetime can lead to geometric connectedness between them~\cite{Ryu:2006bv,VanRaamsdonk:2009ar,VanRaamsdonk:2010pw,VanRaamsdonk:2018zws,Bianchi:2012ev,Caputa:2014eta}. More generally, it has led to the speculation that gravitational theories have an underlying quantum informatic description (see~\cite{Rangamani:2016dms,Faulkner:2022mlp} and references therein).

The oft-cited example of this is a special entangled state, the so-called \textit{thermo-field double} (TFD), which describes a double-sided eternal black hole in AdS space, each side with its own asymptotic region~\cite{Maldacena:2001kr}. This paradigm, commonly referred to as ``ER=EPR'', represents the fact that entangled Einstein-Podolsky-Rosen (EPR) states create an Einstein-Rosen (ER) bridge between them~\cite{Maldacena:2013xja}. The formula due to Ryu and Takanayagi~\cite{Ryu:2006bv} (and refined in~\cite{Hubeny:2007xt}) has put real meat to the above slogan by relating the entanglement entropy between quantum degrees of freedom on the AdS boundary to geometric objects (extremal surfaces) in the dynamical bulk spacetime. For the above double-sided BH, it relates the entanglement entropy to the (minimal) cross-sectional area of the ER bridge. Although this is used as a prototype to explain how spacetime glue emerges from entanglement patterns, it was soon realised that `entanglement is not enough'~\cite{Susskind:2014moa} to get a full picture of the region behind BH horizon. In particular, although the entanglement entropy of each of the subsystem CFTs saturates as they thermalise, the volume of this region keeps growing past the scrambling time~\cite{Susskind:2014rva}. Subsequently, it was proposed that the volume of the ER bridge, joining the two entangled regions outside the BH horizons on either side, is dual to the `quantum computational complexity' of the boundary theory~\cite{Stanford:2014jda}. Through many such holographic conjectures, which have since been proposed~\cite{Brown:2015bva,Couch:2016exn, Chapman:2016hwi, Brown:2017jil,Carmi:2017jqz, Bernamonti:2019zyy}, what has been firmly established is the crucial role `quantum complexity' \cite{Chapman:2021jbh} plays in our understanding of {\it spacetime emergence}.
 
Although ER=EPR has now been extensively studied for some time, our best-understood examples all come from AdS/CFT which has been tested in multiple string theory examples. However, observations tell us that our own universe is accelerating and thus is an asymptotically de Sitter (dS) space~\cite{SupernovaSearchTeam:1998fmf}. Moreover, our best theory of the early universe -- inflation -- also requires a phase of (quasi-)dS expansion~\cite{Planck:2018vyg}. Thus, it behoves us to try to understand these ideas relating the emergence of spacetime from entanglement for a dS universe. In the absence of reliable dS constructions in String Theory to guide us, it is only prudent that we try to understand ER=EPR in dS space within a bottom-up approach.

\begin{figure*}[ht]
\centering

\begin{tikzpicture}

    \begin{scope}[scale=3,xshift=-2.5cm]
    \def\Nlines{4} 
    \coordinate (O) at ( 1, 0); 
    \coordinate (S) at ( 1,-1); 
    \coordinate (N) at ( 1, 1); 
    \coordinate (W) at (0, 0); 
    \coordinate (E) at ( 2, 0); 

    \draw[thick] (N) -- (E) -- (S) -- (W) -- cycle;

    \node[] at (1,1.22) {$CFT_L \otimes CFT_R$};
    \draw[thick,yellow!80!orange,line width=3] (0,1) -- (2,1);

    \foreach \i [evaluate={\c=\i/(\Nlines+1); \cs=sin(90*\c);}] in {1,...,\Nlines}{
      \draw[world line,samples=20,smooth,variable=\x,domain=1:2] 
        plot(\x-1,{1-kruskal(\x*pi/4,\cs)});
      \draw[world line t,samples=20,smooth,variable=\y,domain=0:1] 
        plot({ kruskal(\y*pi/4,\cs)},\y);
      \draw[world line,samples=20,smooth,variable=\x,domain=2:3] 
        plot(\x-1,{1+kruskal(\x*pi/4,\cs)});
      \draw[world line t,samples=20,smooth,variable=\y,domain=0:1] 
        plot({2-kruskal(\y*pi/4,\cs)},\y);
    }
    \draw[thick] (2,1) -- (2,-1) -- (0,-1) -- (0,1);

    \foreach \i [evaluate={\c=\i/(\Nlines+1); \cs=sin(90*\c);}] in {1}{
      \draw[world line thick,samples=20,smooth,variable=\x,domain=1:2] 
        plot(\x-1,{1-kruskal(\x*pi/4,\cs)});
      \draw[world line thick,samples=20,smooth,variable=\x,domain=2:3] 
        plot(\x-1,{1+kruskal(\x*pi/4,\cs)});
    }
    \node[fill=white, inner sep=2] at (O) {\large{\textbf{C}}};
    \node[fill=white, inner sep=2, rotate=90] at (2.1,0) {\large{\textbf{South Pole}}};
    \node[fill=white, inner sep=2, rotate=90] at (-0.1,0) {\large{\textbf{North Pole}}};
    \node[fill=white, inner sep=2] at (1.75,0.7) {{\textbf{R}}};
    \node[fill=white, inner sep=2] at (0.25,0.7) {{\textbf{L}}};

    \node[] at (1.04,1.07) {$\mathcal{I}^+$};
    \node[] at (1.04,-1.07) {$\mathcal{I}^-$};
    \end{scope}

    \begin{scope}[scale=1.4,yshift=-2cm]
    \begin{axis}[
          hide axis,
          view={25}{5},
          domain=0:2*pi,
          samples=100,
          axis equal,
      ]
      \addplot3 [
        domain=0:2*pi,
        samples=100,
        samples y=0,
        thick,
        color=yellow!80!orange, 
        fill=yellow!80!orange, 
        fill opacity=0.3 
      ] (
        {cos(deg(x))}, 
        {sin(deg(x))}, 
        {0}            
      );

      \draw[thick] (0.891,0.454,0) -- (0.891,0.454,-2);
      \draw[thick] (-0.891,-0.454,0) -- (-0.891,-0.454,-2);

      \addplot3 [
        domain=0:2*pi,
        samples=200,
        samples y=0,
        color=violet,
        dashed,
        dash pattern=on 1pt off 1pt, 
        fill=violet, 
        fill opacity=0.5 
      ] (
        {0},                
        {0.08*cos(deg(x))}, 
        {0.08*sin(deg(x))}  
      );

      \path[draw=blue!70!black, thick] (-0.891007,-0.45399,-0.583) arc[start angle=-255, end angle=-269.7, x radius=10cm, y radius=50cm];
      \path[draw=blue!70!black, thick] (-0.891007,-0.45399,-0.583) arc[start angle=270, end angle=285, x radius=10cm, y radius=38cm];

      \path[draw=blue!70!black, thick] (0.891007,0.45399,-0.589) arc[start angle=75, end angle=89.7, x radius=10cm, y radius=50cm];
      \path[draw=blue!70!black, thick] (0.891007,0.45399,-0.589) arc[start angle=-90, end angle=-105, x radius=10cm, y radius=38cm];

      \draw[->] (0.3,0,0.2) -- (0,0,0) node[pos=0,right] {$S^2$};

      \draw[<->] (0.3,0.3,-0.35) -- (0,0,-0.5) -- (-0.3,-0.3,-0.35) node[pos=0,below] {$H^3$};
    \end{axis}

    \begin{scope}[xshift=3.9cm, yshift=3.05cm]
      \foreach \i in {1,2,3,...,10} {
        \pgfmathsetmacro{\angleA}{-255 - \i*1.5} 
        \pgfmathsetmacro{\angleB}{270 + \i*1.5} 
        \pgfmathsetmacro{\xA}{-0.891007 + 10 * (cos(-255)-cos(\angleA))} 
        \pgfmathsetmacro{\yA}{-0.45399 + 50 * (sin(-255)-sin(\angleA))} 
        \pgfmathsetmacro{\xB}{-0.891007 + 10 * (cos(270)-cos(\angleB))}
        \pgfmathsetmacro{\yB}{-0.45399 + 38 * (sin(270)-sin(\angleB))}
        \pgfmathsetmacro{\zA}{-0.583} 
        \pgfmathsetmacro{\zB}{-0.583}

        \draw[red!70!black, thick] ({0.8*\xA}, {0.8*\yA}, {0.8*\zA}) -- ({0.8*\xB}, {0.8*\yB}, {0.8*\zB});
      }
    \end{scope}
    \begin{scope}[xshift=2.6cm, yshift=2.305cm]
      \foreach \i in {1,2,3,...,10} {
        \pgfmathsetmacro{\angleA}{75 + \i*1.5} 
        \pgfmathsetmacro{\angleB}{-90 - \i*1.5} 
        \pgfmathsetmacro{\xA}{0.891007 + 10 * (cos(75)-cos(\angleA))} 
        \pgfmathsetmacro{\yA}{0.45399 + 50 * (sin(75)-sin(\angleA))} 
        \pgfmathsetmacro{\xB}{0.891007 + 10 * (cos(-90)-cos(\angleB))}
        \pgfmathsetmacro{\yB}{0.45399 + 38 * (sin(-90)-sin(\angleB))}
        \pgfmathsetmacro{\zA}{-0.583} 
        \pgfmathsetmacro{\zB}{-0.583}

        \draw[red!70!black, thick] ({0.8*\xA}, {0.8*\yA}, {0.8*\zA}) to ({0.8*\xB}, {0.8*\yB}, {0.8*\zB});
      }
    \end{scope}
    \end{scope}

    \begin{scope}[scale=2,xshift=2cm,yshift=-2.3cm]
    \draw[<->, thick,orange,in=90,out=90] (-3.3, 3.85) to node[pos=0.6,above] {$S^3$}  (-1., 3.02) ;
    \end{scope}
\end{tikzpicture}
\caption{\textbf{(a)} Penrose diagram of dS$_4$ in hyperbolic slicing. The bold blue line represents a typical spacelike hypersurface and the $L$ and $R$ regions are causally disconnected form each other. \textbf{(b)} The yellow-shaded region represents the $S^3$ corresponding to the future conformal boundary, the purple region shows the equatorial $S^2$ in $\mathcal{I}^+$, and the region enclosed by the blue lines represent the volumes of the spacelike hyperboloids.}\label{fig:dS hyper}
\end{figure*}
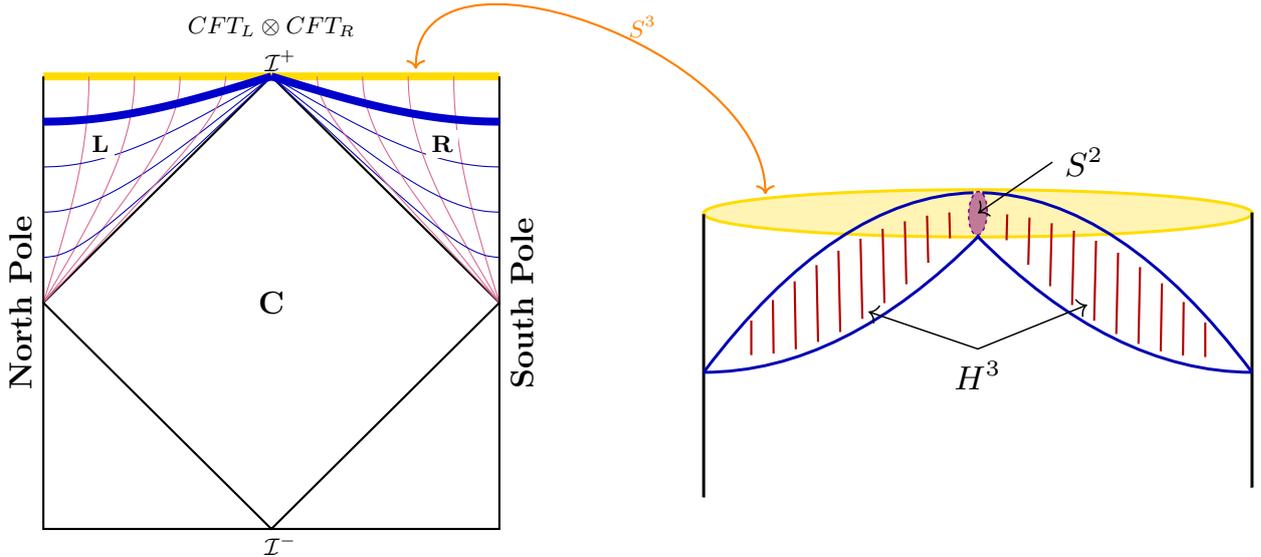

In this vein, it has recently been pointed out that there is a remarkable similarity in the Penrose diagram of full dS spacetime with that of the extended AdS-Schwarzschild spacetime. This led to the conjecture that, unlike AdS/CFT, dS is not dual to a single CFT, rather that the dS spacetime \textit{by itself} emerges from quantum entanglement between two CFTs living on the future conformal boundary $\mathcal{I}^+$~\cite{Cotler:2023xku} (Fig.~\ref{fig:dS hyper}). In this letter, we try to push this line of thought further by asking how ``difficult'' is it to prepare such an entangled state, namely the Bunch-Davies or the Euclidean vacuum? More explicitly, assuming that such a picture of \textit{cosmic ER=EPR} holds in dS/CFT~\cite{Strominger:2001pn}, we quantify the complexity of producing such an entangled TFD state starting from the product state of the two vacua of the boundary CFTs. However, instead of using any holographic dictionary, we will compute the circuit complexity of a free, massive scalar field in dS to show that the entangled Euclidean vacuum is indeed a finite-complexity Hadamard state in the Hilbert space of the theory. (This then naturally generalizes to the case of the graviton vacuum in dS since our findings hold for the massless case.) Moreover, we find some universal properties of this complexity (for the TFD state) and that it agrees with general expectations from the above-mentioned holographic complexity conjectures. 

\noindent {\bf Quantum complexity \`a la Nielsen:} 
For the purposes of this paper, computational or circuit complexity will quantify the difficulty of preparing an entangled (target) state when starting out with some simple enough product (reference) state, given a set of elementary operations/gates. Nielsen geometrized this problem~\cite{Nielsen:2005mkt} by showing that complexity can be measured by geodesics on the group manifold of unitary quantum operations starting out from the identity to an operator that maps reference to target state. Of course, there exist many such operators and the complexity of the target state is then defined as minimising over all such geodesics ending at suitable operators. There exists multiple cost functions which assign costs to the gates required to get to the target state from the reference state. However,~\cite{Jefferson:2017sdb} was showed that we can narrow down to a family of cost functions such that they reproduce the UV-divergence expected from the holographic conjectures for the Minkowski vacuum:
\begin{equation}
\mathcal{D}_\kappa=\int_0^1 {\rm d} s\sum \,\abs{Y^{I}(s)}^\kappa\,,\label{eq:Dalpha}
\end{equation}
where $\mathcal{D}$ is the circuit depth and the control functions $Y^{I}$ determine which gates are used at ``time" $s$. In the geometric language applied to the space of unitary operators, they represent the ``distance'' between the reference and target states and the trajectory, respectively, given some choice of  $\kappa \geq 1 \in \mathbb{R}$. Since the UV-divergence for holographic complexity of dS space has a similar behaviour~\cite{Reynolds:2017lwq}, we assume that the metric on the space of unitaries given by~\eqref{eq:Dalpha} remains reliable, with further justification given below for picking the particular choice of $\kappa=2$.

We will be using a method, developed in~\cite{Hackl:2020ken}, called the covariance matrix method to calculate the circuit complexity. It has already been utilised in the context of Minkowski TFDs~\cite{Chapman:2018hou,Doroudiani:2019llj}, free fermions~\cite{Hackl:2018ptj}, quantum quenches~\cite{Mitra_2018,Ali:2018aon} and inflationary cosmology~\cite{Bhattacharyya:2020rpy,Bhattacharyya:2024duw}. The reason why we are able to directly use this method in our approach is that the reference state, considered to be the product state of the two vacua of the boundary CFTs, as well as the target state, which is the Euclidean vacuum, are both \textit{Gaussian}. The deep physical reason behind this is that the Euclidean vacuum can be realised as a specific entangled state (the TFD state) over the CFT$_3$ vacua. 

The covariance matrix encodes all information of a Gaussian state, as it uniquely parametrizes the wave function up to an unphysical overall complex phase. For a system characterized by $N$ pairs of quadrature field operators\footnote{The quadrature field operators satisfy the standard canonical commutation relations.} $\hat\xi^a\equiv\left(\hat{x}_1, \cdots, \hat{x}_N, \hat{p}_1, \cdots, \hat{p}_N\right)$, the covariance matrix $G$ is defined as
\begin{equation}
G^{ab}=\bra{\psi}\hat\xi^{a}\hat\xi^{b}+\hat\xi^{b}\hat\xi^{a}\ket{\psi} \label{eq:cov mat}\,,
\end{equation}
where $a,b$ labels the indices of the matrix and $\ket{\psi}$ is the corresponding Gaussian state.

The set of unitary operations mapping Gaussian states onto Gaussian states form a projective representation of the Lie group $\mathrm{Sp}(2N,\mathbb{R})$. 
In Nielsen's approach, we can thus compute the geometric complexity as the shortest distance in the manifold $\mathrm{Sp}(2N,\mathbb{R})$, equipped with an appropriately chosen measure of distance. Following~\cite{Hackl:2018ptj,Chapman:2018hou}, we choose a natural right-invariant metric defined in terms of the reference covariance matrix, directly related to the cost function~\eqref{eq:Dalpha} with $\kappa=2$. This choice of $\kappa$ is also natural as it emulates finding the (squared) geodesic in a Riemannian manifold and, as pointed out by~\cite{Jefferson:2017sdb,Chapman:2018hou}, it is invariant under basis changes in the classical phase space of $\hat{\xi}^a$. With these ingredients, the quantum complexity is given by~\cite{Hackl:2020ken}
\begin{equation}
    \mathcal{C}_{\kappa=2}(G_{r},G_{t})=\frac{1}{8}\mathrm{Tr}\left[\mathrm{log}^{2}(G_{t}G_{r}^{-1})\right]\,, \label{eq: cov comp}
\end{equation}
with $G_{t}$ and $G_{r}$ being the target and reference covariance matrices, respectively.

\noindent {\bf de Sitter space in hyperbolic slicing:} Instead of the quasinormal mode (QNM) basis~\cite{Jafferis:2013qia,Cotler:2023xku}, we will employ the hyperbolic slicing\footnote{The hyperbolic slicing simplifies our computations. However, our main result of the finiteness of the complexity will, of course, also hold in the QNM basis \cite{us}.} of dS space (Fig~\ref{fig:dS hyper}), as was used in~\cite{Maldacena:2012xp} to compute the entanglement entropy corresponding to long-range interactions in dS. In this slicing, the metric in the left ($L$) and right ($R$) patches is given by:
\begin{equation}
{\rm d} s_\sigma^2 = H^{-2} \left[-{\rm d} t_\sigma^2 + \sinh ^2 t_\sigma \left({\rm d} r_\sigma^2 + \sinh ^2 r_\sigma\, {\rm d} \Omega_2^2\right)\right]\,,
\end{equation}
where $\sigma$ denotes $L$ or $R$ and $H^{-1}$ is the Hubble horizon.


Following~\cite{Sasaki:1994yt}. we will study the action of a free massive scalar field in this open chart of dS, namely:
\begin{equation}
S=\int {\rm d}^4 x \sqrt{-g}\left[-\frac{1}{2} g^{\mu \nu} \partial_\mu \phi \partial_\nu \phi-\frac{m^2}{2} \phi^2\right]\,.
\end{equation}
The scalar field can be expanded in a basis characterised by the Euclidean vacuum $\ket{BD}$, which we define by the ladder operators $\hat{a}$ and $\hat{a}^{\dagger}$. However, in order to analyse the behaviour in the $L$ and $R$ disconnected regions, we are going to use a different positive frequency mode expansions in each of these regions. Defining the corresponding vacua as $\ket{0}_L$ and $\ket{0}_R$ respectively, the resulting field expansion, in terms of creation and annihilation operators $\hat{b}_{L,R}$ and $\hat{b}^{\dagger}_{L,R}$, can be written as \cite{Sasaki:1994yt, Maldacena:2012xp}
\begin{equation}
    \hat{\phi}_{\sigma}\!=\!\int {\rm d}p\, \tfrac{1}{a(t_\sigma)\, N_{p} }\sum_{l,n}\left(\hat{b}_{\sigma pln}\, P_{\sigma p}\,Y_{\sigma pln} + \hat{b}^{\dagger}_{\sigma pln}\,\bar{P}_{\sigma p}\, \bar{Y}_{\sigma pln}\right),
\end{equation}
where $(p, l, n)$ are quantum numbers labelling different modes and $\sigma$ refers to the $L$ or $R$ patch, $a(t_\sigma)=H^{-1}\mathrm{sinh}(t_\sigma)$ is the scale factor, and $N_{p}$ is some normalisation factor. The mode functions $P_{\sigma p} := P_{\nu-1 / 2}^{i p}\left(\cosh t_{\sigma}\right)$ are also labelled by the mass parameter, $\nu=\sqrt{9/4- m^2/H^2}$. Finally, $Y_{\sigma pln}$ denotes the eigenstates of the Laplacian operator on a $3$-dimensional unit hyperboloid, \textit{i.e.,} $-\mathbf{L}_{\mathbf{H}^{3}}^2 Y_{pln}(r, \Omega)=\left(1+p^2\right) Y_{pln}(r, \Omega)$.

To summarise, we have two different sets of states $\ket{0}_\sigma$ and $\ket{BD}$ characterised by
\begin{eqnarray}
	\hat{b}_{\sigma pln}\ket{0}_\sigma=0\qquad\text{and}\qquad \hat{a}_{\pm p l n} \ket{BD} =0 \,,
\end{eqnarray}
where $\pm$ correspond to the two different solutions. The fact that $\hat{\phi}$ can be expanded in two different bases implies that $\{\hat{a},\hat{a}^{\dagger}\}$ and $\{\hat{b}, \hat{b}^{\dagger}\}$ must be related by a Bogoliubov transformation, so the Hilbert spaces associated with $L$ and $R$ together make up the full Hilbert space of fields in dS (see~\cite{Maldacena:2012xp, Sasaki:1994yt} for details). This relation allows us to write the Euclidean vacuum, mode by mode, in terms of the $\ket{\emptyset}:=\ket{0}_L \otimes \ket{0}_R$ vacuum\footnote{The contribution from the central region is negligible at the boundary CFTs.} as
\begin{equation}\label{BDtoLR}
	\ket{BD} =[\mathrm{det}(\mathbb{I}-M^{\dagger}M)]^{\frac{1}{4}}e^{\frac{1}{2} \sum_{i, j=R, L} M_{i j} \hat{b}_i^{\dagger} \hat{b}_j^{\dagger}}\ket{\emptyset}\,,
\end{equation}
with
\begin{equation}
	M =e^{i \theta} \frac{\sqrt{2} e^{-p \pi}}{\sqrt{\cosh 2 \pi p+\cos 2 \pi \nu}}\left(\begin{array}{cc} \label{eq: mij}
		\cos \pi \nu & i \sinh p \pi \\
		i \sinh p \pi & \cos \pi \nu
	\end{array}\right)\,.
\end{equation}
Note that the mass enters this transformation through $\nu$ and $\theta$ is an unimportant phase parameter that can be reabsorbed into $b^{\dagger}$~\cite{Maldacena:2012xp}.

With the relation~\eqref{BDtoLR} in mind, we will take the $\ket{\emptyset}$ as our reference state and $\ket{BD}$ as the target state to compute the complexity, and elaborate on its connection to cosmic ER=EPR later on.

\noindent {\bf Complexity of formation of the Euclidean vacuum:} Let us now compute the quantum complexity for the Euclidean vacuum in dS
adopting the covariance matrix method. We call this the complexity of formation of the Euclidean vacuum, following the nomenclature in ~\cite{Chapman:2016hwi}, as we are starting with the $\ket{\emptyset}$ state from the beginning and our goal is to extract the complexity that comes from the entanglement between these two regions. To fully describe the Euclidean vacuum in dS space, we choose our array of operators as $\hat{\xi}^a\equiv(\hat{x}_{L}^{I},\hat{x}_{R}^{I}, \hat{p}_{L}^{I},\hat{p}_{R}^{I})$ with $\hat{x}=\tfrac{\hat{b}+\hat{b}^\dagger}{\sqrt{2}}$ and $\hat{p}=\frac{\hat{b}-\hat{b}^\dagger}{\sqrt{2}i}$, where $I$ collectively denotes the quantum numbers $(p,l,n)$.

The reference and target covariance matrices $G_r$ and $G_t$, respectively, are computed based on~\eqref{eq:cov mat} with respect to  $\ket{\emptyset}$ and $\ket{BD}$.
We can always perform a change of basis such that $G_r$ is the identity matrix. Through some algebra and block-decomposition~\cite{Windt:2020tra}, the elements of $G_t$ can be written in terms of~\eqref{eq: mij} as
\begin{eqnarray}
    G_{t}=\oplus_{I}\left(\begin{array}{ll}
 G_{1}\; G_{2} \\
 G_{3}\; G_{4}\\
\end{array}\right)_{I} \ ,
\end{eqnarray}
where
\begin{align}
G_1&=\operatorname{Re}\left[\left(\mathbb{I}_2+2 M+M^{\dagger} M\right)\left(\mathbb{I}_2-M^{\dagger} M\right)^{-1}\right],\\
G_2&=\operatorname{Im}\left[\left(\mathbb{I}_2+2 M+M^{\dagger} M\right)\left(\mathbb{I}_2-M^{\dagger} M\right)^{-1}\right],\\
G_3&=\operatorname{Im}\left[\left(-\mathbb{I}_2+2 M-M^{\dagger} M\right)\left(\mathbb{I}_2-M^{\dagger} M\right)^{-1}\right],\\
G_4&=\operatorname{Re}\left[\left(\mathbb{I}_2-2 M+M^{\dagger} M\right)\left(\mathbb{I}_2-M^{\dagger} M\right)^{-1}\right] \  .
\end{align}
Given the explicit form of the covariance matrices, we can compute the eigenvalues of $G_tG_r^{-1}$ to consist of quadruples $\tanh(\tfrac{p\pi}{2})$, so the total complexity w.r.t. $\kappa=2$ is
\begin{align}\label{Main_result}
        \mathcal{C}_{\kappa=2}(G_{r},G_{t})&=\frac{1}{8}\mathrm{Tr}\left[\mathrm{log}^{2}(G_{t}G_{r}^{-1})\right]\\
        &=\frac{V_{\mathbf{H}^{3}}}{4\pi^2}\int {\rm{d}} p\, p^{2}\log^{2}\left[\tanh(\frac{p\pi}{2})\right],
\end{align}
where we integrated over momentum space with the volume form $\frac{V_{\mathbf{H}^{3}}}{(2\pi)^3}\int {\rm{d}}^3p$, with $V_{\mathbf{H}^{3}}$ denoting the volume of the hyperboloid; $\frac{p^{2}}{2\pi^{2}}$ is the measure of the density of states.
At leading order in $p$, the integrand $p^2\mathrm{log}^{2}(\tanh{\frac{p\pi}{2}})$ behaves like $p^2 \log^2(\tfrac{p\pi}{2})$ as $p\to 0$ and like $4p^2e^{-2 \pi p}$ as $p\to\infty$ and is thus integrable on the positive real line and yields $c=0.03598\dots$ when evaluated numerically. We emphasize that the momentum integral \textit{does not have any of the divergences} encountered in flat-space QFT\footnote{There is a trivial divergence in the pre-factor due to the volume of the hyperboloid $V_{{\bf H}_3}$ since we take the entangling surface at the future conformal boundary \cite{Maldacena:2012xp}.}.

\noindent{\bf Evidence for \textit{cosmic ER=EPR}:} The  dS/CFT conjecture posits that quantum gravity in the bulk can be understood by analysing the holographic dual CFT living on $\mathcal{I}^+$. The idea behind the \textit{cosmic ER=EPR} slogan was that bulk states in dS are dual to entangled TFD states in the two copies of the boundary CFT Hilbert space \textit{i.e.} $\mathcal{H}_{\text{dS}_4} \simeq \mathcal{H}_{\text{CFT}_3} \otimes \mathcal{H}_{\text{CFT}_3}$. 

To be more precise, note that states in the dual CFT\footnote{Insert an operator at the north pole, and define a state corresponding to it as a path integral with boundary conditions specified on the equatorial $S^2$. Such an operator/state correspondence has been explicitly verified for higher spin gravity in~\cite{Ng:2012xp}.} can be defined on the equatorial $S^2$ in the $S^3$ at $\mathcal{I}^+$ (see Fig.~\ref{fig:dS hyper}). However, in the global slicing of dS, the bulk states live on the spacelike $S^3$ hypersurfaces which cannot be identified with any unique $S^2$ on $\mathcal{I}^+$~\cite{Ng:2012xp}, and thus one should look at spacelike hypersurfaces that end in the $S^2$ at $\mathcal{I}^+$. Since there are two such hypersurfaces (Fig.~\ref{fig:dS hyper}), the Hilbert space of dS quantum gravity can be thought of as a tensor product of two copies of the CFT$_3$ on $\mathcal{I}^+$. Using a QNM basis, it was also shown that the ground state for a massive scalar field in dS, \textit{i.e.} the Euclidean vacuum, is precisely a TFD state in the product Hilbert space of the boundary CFTs~\cite{Cotler:2023xku}. 

Since we have used a hyperbolic slicing~\cite{Sasaki:1994yt, Maldacena:2012xp} instead of the QNM basis used in~\cite{Cotler:2023xku} to propose \textit{cosmic ER=EPR}, we first show that the Euclidean vacuum can still be written as a TFD state over the CFT vacua. First, note that a TFD state, constructed from the energy eigenstates $(\ket{E_\alpha}_{L},\ket{E_\alpha}_{R})$ corresponding to each copy of the CFT$_3$, has the general form:
\begin{equation}
\ket{TFD}=\frac{1}{\sqrt{Z(\beta)}} \sum_\alpha e^{-\beta E_\alpha / 2}\ket{E_\alpha}_L\ket{E_\alpha}_R\,. \label{eq:TFD}
\end{equation} 
Since the Euclidean vacuum~\eqref{BDtoLR} can be rewritten in a different frame (that does not mix $L$ and $R$ modes~\cite{Maldacena:2012xp}) as
\begin{align}
\ket{BD}&=\sqrt{1-\abs{\gamma}^2}\ e^{\frac{1}{2}\gamma \hat{c}_L^{\dagger} \hat{c}_R^{\dagger}}\ket{0}_{L^\prime} \otimes\ket{0}_{R^\prime}\\
&= \sqrt{1-\abs{\gamma}^2} \ \sum_\alpha\abs{\gamma}^\alpha \ket{E_\alpha}_{L^\prime} \otimes \ket{E_\alpha}_{R^\prime}\ ,
\end{align}
we immediately see that it is identical to~\eqref{eq:TFD} from the second equality, on identifying that $\abs{\gamma}^\alpha  = \exp(-\beta E_{\alpha}/2)$, with $Z(\beta)=\frac{1}{1-\abs{\gamma}^2}$. It was already shown that the Euclidean vacuum can be written as a TFD state in the bulk, with respect to the $\ket{0}_L$ and $\ket{0}_R$ vacua, in~\cite{Maldacena:2012xp} with the natural choice for the (inverse of the) Gibbons-Hawking temperature as $\beta = H/(2\pi)$. The thermal reduced density matrix for each of the patches can, in turn, be written as $\rho_{\rm red}=e^{-\beta \mathbb{H}_{\mathrm{ent}}}$, where $\mathbb{H}_{\mathrm{ent}}$ denotes the entangling Hamiltonian. We identify each of the $L$ and $R$ vacua with the vacua of the (dual) boundary CFTs in the spirit of \textit{cosmic ER=EPR}. 

This discussion shows that the hyperbolic slicing naturally suggests that the two CFTs are localised on the two hemispheres of $S^3$ on $\mathcal{I}^+$ and the entanglement between them leads to the emergence of the bulk Euclidean state for a massive scalar field. Let us discuss the implications of our main result for the complexity of this state. Note that our reference state $\ket{\emptyset}$ (which, in the dual theory, is the product state between the two CFT$_3$ vacua) does not live in $\mathcal{H}_{{\rm dS}_4}$ since it does not have zero charge under the dS$_4$ isometries.  Our result shows that the complexity density (defined as complexity per unit volume) to form the Euclidean vacuum in dS$_4$ is \textit{finite both in the UV and in the IR}, and there is no cutoff necessary to regularise the momentum integral. This signals an absence in arbitrary long-distance entanglement responsible for forming this TFD state from the two local vacua. Contrast our result with the usual UV divergence of the complexity of the Minkowski vacuum, when starting out with ultra-local vacua on a lattice. This is related to the fact that both $\ket{\emptyset}$ and $\ket{BD}$ live in the same unitary representation of our field operator algebra\footnote{A straightforward calculation shows that the expected particle number w.r.t. $\hat{n}_{+pln}=\hat{a}^\dagger_{+pln}\hat{a}_{+pln}$ in the state $\ket{\emptyset}$ is finite even after summing/integrating over $pln$, thus establishing that both states live in the same representation~\cite{agullo2015unitarity}.}, while the locally unentangled field vacuum in Minkowski space and the Poincar\'e vacuum do not. Thus, the finite complexity implies that we can think of the Euclidean vacuum as living in the $\mathcal{H}_{\text{CFT}_3} \otimes \mathcal{H}_{\text{CFT}_3}$ Hilbert space, thereby further supporting the \textit{cosmic ER=EPR} slogan. The physical consequence of this is similar to the ER=EPR story in the case of the double-sided BH: The entanglement between two causally-disconnected regions (inside the future light cones of the two inertial observers at the North and South Poles) give rise to entire bulk states in full dS$_4$.

\noindent{\bf Discussion:} It has been pointed out in~\cite{Reynolds:2017lwq} that the holographic conjectures imply that the complexity density in dS-invariant states must be time-independent, as opposed to the linear behaviour expected for black holes. Our result not only verifies this, but shows that the complexity density, which corresponds to the momentum integral in~\eqref{Main_result}, is a universal numerical factor.

This provides a segue into another peculiar feature of the resulting complexity. It is independent of the parameters in the Lagrangian, namely the mass of the scalar field, and only depends on the geometry of the background through the dependence of the volume of the hyperboloid on $H$. The nontrivial integral in~\eqref{Main_result} is simply a finite numerical factor independent of $\nu$ even though the transformation $M_{ij}$ depends on it. Since the causal structure is very similar to how the global Minkowski vacuum looks as an entangled state over the Rindler vacua, the complexity in that  case reveals a similar universal number independent of the mass of the scalar field and dependent only on the acceleration of the observer \cite{us}. This seems to be hinting at a new ``Third Law'' of complexity: For a spacetime with horizons such that a global state emerges from the entanglement between two causally disconnected, local vacua, the resulting complexity density is characterised by a universal numerical factor independent of the details of the field theory under consideration (such as the mass of the field). The only dependence that the complexity of such a state has comes from a prefactor dependent on the geometry (and dimension) of the resulting spacetime. This is a hint that \textit{cosmic ER=EPR} goes beyond QFT in dS and indeed holds for quantum gravity in dS space. Moreover, although we demonstrate our results using the TFD state for a massive scalar field in dS, as was done in \cite{Cotler:2023xku}, the same results would hold, at the very least, for perturbative quantum gravity. Gravitons are metric fluctuations, $h_{\mu\nu}(x,t) := g_{\mu\nu}(x,t) - \eta_{\mu\nu}^{\rm(dS)}(t)$, the action for which is the same as that for (two-copies of) a massless scalar field in dS. Since our result for finite complexity is independent of the mass parameter, it holds for dS gravitons.

Finally, let us emphasize the main \textit{physical} implication of our result. Going beyond {\it cosmic ER=EPR}, we have discovered a new feature for spacetimes with horizons. There is a natural choice for the reference state as a product state of the vacua of two causally-disconnected regions for which the global vacuum can be written as a special entangled (TFD) state over this reference state. It turns out that the complexity, corresponding to the long-distance entanglement required to describe the region ``outside'' the horizons of the local regions, is both \textit{UV- and IR-finite}. Since our result concerns complexity of constructing TFDs from two-copies of local Hamiltonians \cite{Cottrell:2018ash}, it will find applications more generally for studying thermalization in quantum systems. 


\vspace{2mm}
\noindent{\bf \textit{Acknowledgments}:} SB thanks Sumit Das, Keshav Dasgupta, Jan Pieter, Ashish Shukla and Joan Simon for extensive discussions. SB is supported in part by the Higgs Fellowship and by the STFC Consolidated Grant ``Particle Physics at the Higgs Centre''. LH acknowledges support by the Alexander von Humboldt Foundation, by grant $\#$62312 from the John Templeton Foundation, as part of the \href{https://www.templeton.org/grant/the-quantuminformation-structure-ofspacetime-qiss-second-phase}{‘The Quantum Information Structure of Spacetime’ Project (QISS)}, by grant $\#$63132 from the John Templeton Foundation and an Australian Research Council Australian Discovery Early Career Researcher Award (DECRA) DE230100829 funded by the Australian Government. XL is supported in part by the Program of China Scholarship Council (Grant No. 202208170014). The opinions expressed in this publication are those of the authors and do not necessarily reflect the views of the respective funding organization.

\end{document}